\documentclass[useAMS,twocolumn]{mn2e}
\usepackage{amssymb}
\usepackage{graphicx}

\def\be{\begin{equation}}
\def\ee{\end{equation}}

\catcode`\@=11 
\def\@versim#1#2{\vcenter{\offinterlineskip
        \ialign{$\m@th#1\hfil##\hfil$\crcr#2\crcr\sim\crcr } }}

\usepackage[dvips]{color}

\begin{document}
\title[MTI and MRI in 2-dimensional accretion flow]
{Magnetothermal and magnetorotational instabilities in hot accretion
flows}
\author[Bu, Yuan \& Stone]
{De-Fu Bu$^{1}$\thanks{E-mail:dfbu@shao.ac.cn },
Feng Yuan$^{1}$\thanks{E-mail:fyuan@shao.ac.cn} and James M. Stone$^{2}$\\
$^{1}$Key Laboratory for Research in Galaxies and Cosmology,
Shanghai Astronomical Observatory,\\ Chinese Academy of
Sciences, 80 Nandan Road, Shanghai 200030, China\\
$^{2}$Department of Astrophysical Sciences, Peyton Hall, Ivy Lane,
Princeton, NJ 08544, USA\\}

\maketitle

\date{Accepted . Received ; in original form}

\markboth{}{}
\begin{abstract}
In a hot, dilute, magnetized accretion flow, the electron mean-free
path can be much greater than the Larmor radius, thus thermal
conduction is anisotropic and along magnetic field lines.  In this
case, if the temperature decreases outward, the flow may be subject
to a buoyancy instability (the magnetothermal instability, or MTI).
The MTI amplifies the magnetic field, and aligns field lines with
the radial direction. If the accretion flow is differentially
rotating, the magnetorotational instability (MRI) may also be
present. Using two-dimensional, time-dependent magnetohydrodynamic
simulations, we investigate the interaction between these two
instabilities. We use global simulations that span over two orders
of magnitude in radius, centered on the region around the Bondi
radius where the infall time of gas is longer than the growth time
of both the MTI and MRI.  Significant amplification of the magnetic
field is produced by both instabilities, although we find that the
MTI primarily amplifies the radial component, and the MRI primarily
the toroidal component, of the field, respectively. Most
importantly, we find that if the MTI can amplify the magnetic energy
by a factor $F_t$, and the MRI by a factor $F_r$, then when the MTI
and MRI are both present, the magnetic energy can be amplified by a
factor of $F_t \cdot F_r$. We therefore conclude that amplification
of the magnetic energy by the MTI and MRI operates independently. We
also find that the MTI contributes to the transport of angular
momentum, because radial motions induced by the MTI increase the
Maxwell (by amplifying the magnetic field) and Reynolds stresses.
Finally, we find that thermal conduction decreases the slope of the
radial temperature profile.  The increased temperature near the
Bondi radius decreases the mass accretion rate.
\end{abstract}
\markboth{}{}

\begin{keywords}
accretion, accretion discs -- conduction -- magnetohydrodynamics:
MHD -- ISM: jets and outflow -- black hole physics
\end{keywords}

\section{INTRODUCTION}
Hot, diffuse accretion flows are likely present in low-luminosity active
galactic nuclei (AGN) and the quiescent and hard states of black hole
X-ray binaries (Narayan 2005; Yuan 2007; Narayan \& McClintock 2008).
When the mass accretion rate is very low, the electron mean-free path
will be very large, so the accreting plasma is nearly collisionless.
For example, from $Chandra$ observations of the accretion flow at the
Galactic Center, Sgr A*, the electron mean-free path is estimated to be
$0.02-1.3$ times the Bondi radius ($\sim 10^5-10^6$ Schwarzschild radii).
In such cases, thermal conduction can have significant influence on
the dynamics (Quataert 2004; Johnson \& Quataert 2007), resulting in
the transport of thermal energy from the inner (hotter) to the outer (cooler)
regions. If the energy flux carried by thermal conduction is substantial,
the temperature of the gas in the outer regions can be increased above the
virial temperature. Thus, gas in the outer regions is able to escape from
the gravitational potential of the central black hole and form outflows,
significantly decreasing the mass accretion rate (Tanaka \& Menou 2006;
Johnson \& Quataert 2007).

Time-dependent hydrodynamical simulations of hot accretion flows
show that they are convectively unstable, no matter whether the flow
is radiatively inefficient (Igumenshchev \& Abramowicz 1999; Stone,
Pringle \& Begelman 1999; Igumenshchev \& Abramowicz 2000) or
efficient (Yuan \& Bu 2010).  This is consistent with the prediction
of one-dimensional analytical models of advection-dominated
accretion flows (ADAFs; Narayan \& Yi 1994; 1995).   Convective
instability occurs because the entropy of the flow increases in the
direction of gravity.  However, in a magnetized weakly collisional
accretion flow, in which the electron mean-free path is much greater
than the Larmor radius, thermal conduction is anisotropic and
primarily along magnetic field lines.  In this case, Balbus (2000;
2001) found that the stability criterion for convection is
fundamentally altered: convection occurs when the temperature -- not
the entropy -- increases in the direction of gravity.   Convective
instabilities in this regime are referred to as the magnetothermal
instability (MTI).

In a weakly magnetized, non-rotating atmosphere, local simulations
have demonstrated that the MTI can amplify the magnetic field and
lead to a substantial convective heat flux (Parrish \& Stone 2005;
2007).  Sharma, Quataert \& Stone (2008; hereafter SQS08)
investigated the effects of the MTI on accretion flows in
two-dimensions using global simulations in spherical polar
coordinates. Their simulations focused on the region around the
Bondi radius, where the MTI growth time is shorter than gas infall
time.  They found the MTI can amplify the magnetic field and align
field lines with the direction of the temperature gradient
(radially), consistent with the results obtained by previous local
simulations. The effects of the MTI on the intracluster medium (ICM)
have also been investigated (Parrish, Stone \& Lemaster 2008).  In
that study also, the MTI drives the magnetic field lines to be
radial, leading to a conductive flux which is an appreciable
fraction of the classical Spitzer value. Such a high rates of
thermal conduction make the slope of the temperature profile
flatter.

The studies of the MTI described above all focus on a non-rotating
accretion flow. It is well-known that in a weakly magnetized,
differentially rotating hot accretion flow, the magnetorotational
instability (MRI; Balbus \& Hawley 1991; 1998) will be present, and
MHD turbulence driven by the MRI is thought to be the most promising
mechanism of angular momentum transport. In principle the MTI and
MRI should coexist in such flows. Therefore, an interesting question
for investigation is whether the MTI and MRI interact with one
another, i.e., does one instability enhance or suppress the other,
or do the two instabilities operate entirely independently.  For
example, it is of interest to ask whether convective motions driven
by the MTI can transport angular momentum, and similarly whether
turbulent motions driven by the MRI can transport heat.

Motivated by the question above, we present the results of
two-dimensional magnetohydrodynamical (MHD) simulation of hot
accretion flows with anisotropic thermal conduction and rotation. We
investigate flows which are slowly rotating (not rotationally
supported) at the Bondi radius, where the growth time of both the
MTI and MRI is shorter than the infall time.  This paper is
organized as follows: we introduce our numerical methods and initial
conditions in Section 2, most of our results are presented in
Section 3, while Section 4 is devoted to a summary and discussion.

\section{NUMERICAL SIMULATION}

We adopt two-dimensional, spherical coordinates $(r, \theta)$ and assume
axisymmetry $\partial/\partial \phi \equiv 0$.  We use the Zeus-2D
code (Stone \& Norman 1992a, 1992b) to solve the MHD equations with
anisotropic thermal conduction:
\begin{equation}
\frac{d\rho}{d{t}}+\rho\nabla\cdot\mathbf{v}=0,
\end{equation}
\begin{equation}
\rho\frac{d\mathbf{v}}{dt}= -\nabla p-\rho\nabla\psi
+\frac{1}{4\pi}(\nabla\times\mathbf{B})\times\mathbf{B},
\end{equation}
\begin{equation}
\frac{\partial{\mathbf{B}}}{\partial t}=\nabla\times(\mathbf{v}
\times \mathbf{B}),
\end{equation}
\begin{equation}
\rho\frac{d (e/\rho)}{d t}=-p\nabla\cdot
\mathbf{v}-\nabla\cdot\mathbf{Q},
\end{equation}
\begin{equation}
\mathbf{Q}=-\chi\widehat{\mathbf{b}}(\widehat{\mathbf{b}}\cdot\nabla)T.
\end{equation}
Here $\rho$ is the mass density, $\mathbf{v}$ is the velocity, $p$
is the pressure, $\psi=-GM/r$ is the gravitational potential ($G$ is
the gravitational constant and M is the mass of the central black
hole), $\mathbf{B}$ is the magnetic field, $ e=p/(\gamma-1) $ is the
internal energy ($\gamma$ is the adiabatic index), $\mathbf{Q}$ is
the heat flux along magnetic field lines, $\chi$ is the thermal
diffusivity, $\widehat{\mathbf{b}}$ is the unit vector along
magnetic field and T is the temperature of the gas.

\textbf{For convenience, we adopt $\chi=\kappa p/T$ and assume that
$\kappa=a_c(GMr)^{1/2}$ which has the dimension of a diffusion
coefficient (${\rm cm^2\,s^{-1}}$). }Anisotropic thermal conduction
is implemented using the method based on limiters (Sharma \& Hammett
2007) to guarantee that heat always flows from hot to cold regions.

We initialize our simulations as follows. The radial velocity,
internal energy, and density of the accretion flow are adopted from
the hydrodynamic Bondi solution. For the other two components of the
velocity ($v_{\theta}$ and $v_{\phi}$), we set $v_{\theta}=0$ and
$v_{\phi}=l_0/r$ with $l_0(=const.)$ (a parameter of our
simulations) the specific angular momentum of the accretion flow.
The initial magnetic field is uniform and perpendicular to the
equatorial plane, (i.e. ${\bf B} = B_z$, where $z$ is the vertical
cylindrical coordinate).  We have run four models. The parameters
used in these models are listed in Table 1. In models L1 and L2, we
set $l_0$ to be $0.1$ times the Keplerian angular momentum at the
inner boundary. Thus, in models L1 and L2, the centrifugal barrier
is inside our inner boundary, and rotationally supported region is
never formed in these two models. In models H1 and H2, however, we
set $l_0$ to be $10$ times the Keplerian angular momentum at the
inner boundary. Thus, in models H1 and H2, the centrifugal barrier
is outside our inner boundary, so that a rotationally supported disk
can form, which is subject to the MRI. We set $\beta$ (the ratio of
gas pressure to magnetic pressure) to be $\beta=10^{12}$ at the
outer boundary in our initial conditions. We assume that the
adiabatic index $\gamma=1.5$. We use the Bondi radius,
$r_B=GM/c_\infty^2$ ($c_{\infty}$ is the sound speed at infinity and
we set $c_{\infty}^2=10^{-6}c^2$ where $c$ is the speed of light) to
normalize all of the length scales in our paper. Times reported in
this paper are all in the unit of the inverse rotation frequency at
the outer radius of the domain,
$\Omega_{out}^{-1}=[r_{out}^3/GM]^{1/2}$.

All of our calculations span a domain bounded by an inner boundary at $r_{in}=0.05r_B$ and
an outer boundary at $20r_B$. As in SQS08, we adopt a numerical resolution of
$60\times 44$. The radial grid is logarithmically spaced. A uniform
polar grid extends from $\theta=0$ to $\pi$. At the poles, we use
axisymmetric boundary conditions. In the radial direction, we use
inflow/outflow boundary conditions at both the inner and outer
boundaries. For the temperature at the inner and outer boundaries,
we use outflow boundary conditions.

\section{results}

\subsection{Magnetic field amplification}

A weakly magnetized, conductive, differentially rotating accretion
flow is subject to magnetothermal and magnetorotational
instabilities when the condition,
\begin{equation}
-\frac{1}{\rho}\frac{dp}{dr}\frac{d {\rm ln}T}{dr}+\frac{d
\Omega^2}{d {\rm ln} r}<0,
\end{equation}
is satisfied,
where $\Omega$ is the angular velocity of the accretion flow.
Provided that the magnetic field is weak, the condition for the
existence of MTI alone is
\begin{equation}
-\frac{1}{\rho}\frac{dp}{dr}\frac{d {\rm ln}T}{dr}<0.
\end{equation}
This condition is always satisfied in the initial conditions of
our calculations, thus as long as
the MTI growth time is smaller than the gas
infall time, the MTI will grow.
The condition for the existence of the MRI is
\begin{equation}
\frac{d \Omega^2}{d {\rm ln} r}<0.
\end{equation}
This condition is again always satisfied in the initial conditions
of our calculations.
Defining
\begin{equation}
\gamma_{MTI}^2=\left|-\frac{1}{\rho}\frac{dp}{dr}\frac{d {\rm
ln}T}{dr}\right|,
\end{equation}
\begin{equation}
\gamma_{MRI}^2=\left|\frac{d \Omega^2}{d {\rm ln} r}\right|,
\end{equation}
the MTI and MRI growth timescales are then
$t_{MTI}=\gamma_{MTI}^{-1}$ and $t_{MRI}=\gamma_{MRI}^{-1}$,
respectively. The wavelength of the maximum growth rate for MTI and
MRI are \begin{equation} \lambda_{MTI}=2\pi v_a/\gamma_{MTI}
\end{equation} and
\begin{equation} \lambda_{MRI}=2\pi v_a/\gamma_{MRI}, \end{equation} respectively.
Here $v_a(\equiv\sqrt{B^2/4 \pi \rho})$ is the Alfven speed.

\begin{table}
\footnotesize
\begin{center}
\caption{Parameters for all of our models}
\begin{tabular}{cccc}
\hline \hline
Models & $l_0^\dagger$ & $a_c^\ddagger$ & description  \\
\hline
L1 & $ 0.1 l_{in}$ & $0$  & no MTI, no MRI  \\
L2 & $ 0.1 l_{in}$ & $0.2$ & only MTI \\
H1 & $ 10 l_{in}$ & $0$  &    only MRI\\
H2 & $ 10 l_{in}$ & $0.2$ & with MRI, with MTI \\
\hline
\end{tabular}

$^\dagger$ $l_0$ is the initial angular momentum of the flow.
$l_{in}$ is the Keplerian angular momentum at the inner boundary.\\
$^\ddagger$ $\chi=\kappa p/T$ is the thermal diffusivity and
$\kappa=a_c(GMr)^{1/2}$. \\
\end{center}
\end{table}

We investigate the interplay between the MTI and MRI by examining
the magnetic field amplification by the two instabilities. Note that
in addition to the MTI and MRI, the magnetic field can also be
amplified by the geometric convergence of the accretion flow in
spherical geometry and flux freezing. First, we investigate the
magnetic field amplification by flux freezing by carrying out model
L1. We note that in models L1 and L2, the centrifugal barrier is
inside our inner boundary, thus the timescale for the MRI to grow is
much longer than the gas infall timescale. Furthermore, in model L1
we set the coefficient of the thermal diffusivity $a_c=0$, i.e.,
there is no MTI in model L1.  Therefore, the magnetic field
amplification in model L1 can only be due to flux freezing. The
results below show that the effect of flux freezing on magnetic
field amplification is very small, allowing us to ignore this effect
in the analysis of models L2, H1 and H2.

In model L2, we set the coefficient of the thermal diffusivity
$a_c=0.2$ so that it is unstable to the MTI. In this model, the
magnetic field is amplified only by the MTI. In models H1 and H2,
the centrifugal barrier is outside our inner boundary. The timescale
for the growth of the MRI is much shorter than the gas infall time,
so that the MRI will affect the flow in both of these two models. In
model H1, we set the coefficient of the thermal diffusivity $a_c=0$,
so that the MTI is suppressed and the magnetic field is amplified
only by the MRI. In model H2, we set the coefficient of the thermal
diffusivity $a_c=0.2$, so that both the MTI and MRI coexist, and the
magnetic field is amplified by the nonlinear evolution of both. The
properties of the four models are summarized in Table 1.  The models
demonstrate the interplay between the MTI and MRI. We have also
tried various other values of $a_c$, and find that changing $a_c$ by
a factor of a few does not affect our results, consistent with the
conclusion in SQS08.

\begin{figure}
\begin{center}
\includegraphics[width=8.5cm]{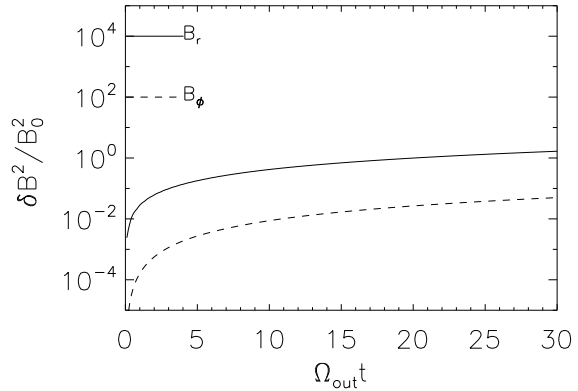}

\end{center}
\caption{Evolution of the volume averaged magnetic energy in both
the $r$ and $\phi$ components of the magnetic field normalized to
the initial magnetic energy in a shell from $r_{\rm out}/40$ to
$r_{\rm out}$ in model L1. In this model, magnetic field is
amplified only by flux freezing.  Since $v_{\theta}\sim 0$, there is
almost no amplification of the $\theta$ component of the field, so
it is not shown. The growth of each component is defined as $\delta
B^2_{r,\theta,\phi} = ({\langle
{B^2}_{r,\theta,\phi}\rangle}_V-B^2_{r,\theta,\phi,0})$, where
$\langle\rangle_V$ denotes volume average and $B_0$ is the initial
field strength.}
\end{figure}

Figures 1-4 show the change of magnetic energy as a function of time
in these four models. The energy is averaged from $r=r_{out}/40$ to
$r_{out}$. The solid, dotted, and dashed lines correspond to the
energy in the $r$, $\theta$, and $\phi$ components of the magnetic
field, respectively.

Figure 1 shows the magnetic energy evolution in model L1. The field is
smoothly amplified by flux freezing. At the end of the simulation,
the field is only amplified by a factor of $2$. Because
$v_{\theta}\sim 0$, there is almost no amplification of the $\theta$
component of the field. From this model, we can see that the effect
of flux freezing on magnetic field amplification is very small.

We now investigate the magnetic field amplification by the MTI
(model L2). The only difference between models L1 and L2 is that in
the latter we include thermal conduction so that the MTI is present.
Fig. 2 shows the evolution of the magnetic energy in model L2, it
shows significantly more amplification of the magnetic field in
comparison to model L1. This amplification is mainly due to motions
driven by the MTI. Initially, the field is rapidly amplified by the
MTI. After time $t=6$, the magnetic field ceases to grow
exponentially, and the growth rate becomes small, for the following
reason. The growth rate of the fastest growing mode of the MTI is
given by $\gamma\simeq |[1/\rho][dp/dr][dlnT/dr]|^{1/2}b_{\theta}$
where $b_{\theta}=B_{\theta}/\textbf{B}$. The growth rate decreases
with time because of the decrease of $b_{\theta}$ with time. In this
model, we use parameters $F_{r,t}$, $F_{\theta,t}$ and $F_{\phi,t}$
to represent the amplification factor of the $r$, $\theta$ and
$\phi$ components of the magnetic field, respectively. We find
\begin{equation}
F_{r,t}=B_{r,f}^2/B_0^2 \sim 100,
\end{equation}
\begin{equation}
F_{\theta,t}=B_{\theta,f}^2/B_0^2 \sim 10,
\end{equation}
\begin{equation}
F_{\phi,t}=B_{\phi,f}^2/B_0^2 \sim 100,
\end{equation}
where $B_{r,f}$, $B_{\theta, f}$ and $B_{\phi,f}$ correspond to the
$r$, $\theta$ and $\phi$ components of the magnetic field at the end
of the simulation, respectively. The radial component of the
magnetic field is amplified by the MTI significantly.  The MTI saturates by
aligning field lines with the temperature gradient (radial
direction), which is consistent with the results of previous work
(Parrish \& Stone 2005; 2007; SQS08). SQS08 found that the magnetic
field amplification by MTI is quasi-linear.  If the magnetic field is
weak initially, MTI can amplify the field by a factor of $\sim
10-30$, independent of the initial field strength. Our calculation
confirms their result.

\begin{figure}
\begin{center}
\includegraphics[width=8.5cm]{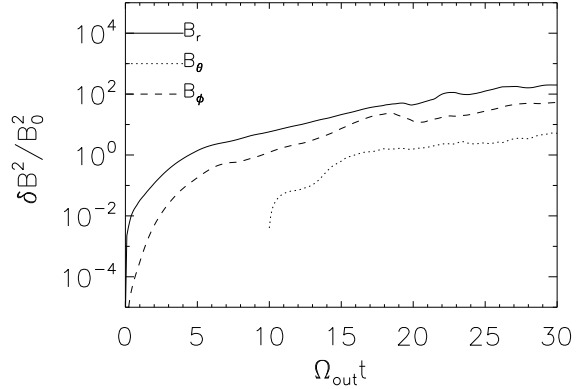}\\

\end{center}
\caption{Change of volume averaged magnetic energy in the $r$, $\theta$
and $\phi$ components of the magnetic field normalized to the initial magnetic energy in
a shell from $r_{\rm out}/40$ to $r_{\rm out}$ as a function of time
in model L2. In this model, the magnetic field is amplified by the MTI. The
growth of each component of the magnetic field is defined as $\delta
B^2_{r,\theta,\phi} = ({\langle
{B^2}_{r,\theta,\phi}\rangle}_V-B^2_{r,\theta,\phi,0})$, where
$\langle\rangle_V$ denotes volume average and $B_0$ is the initial
field strength.}
\end{figure}

\begin{figure}
\begin{center}
\includegraphics[width=8.5cm]{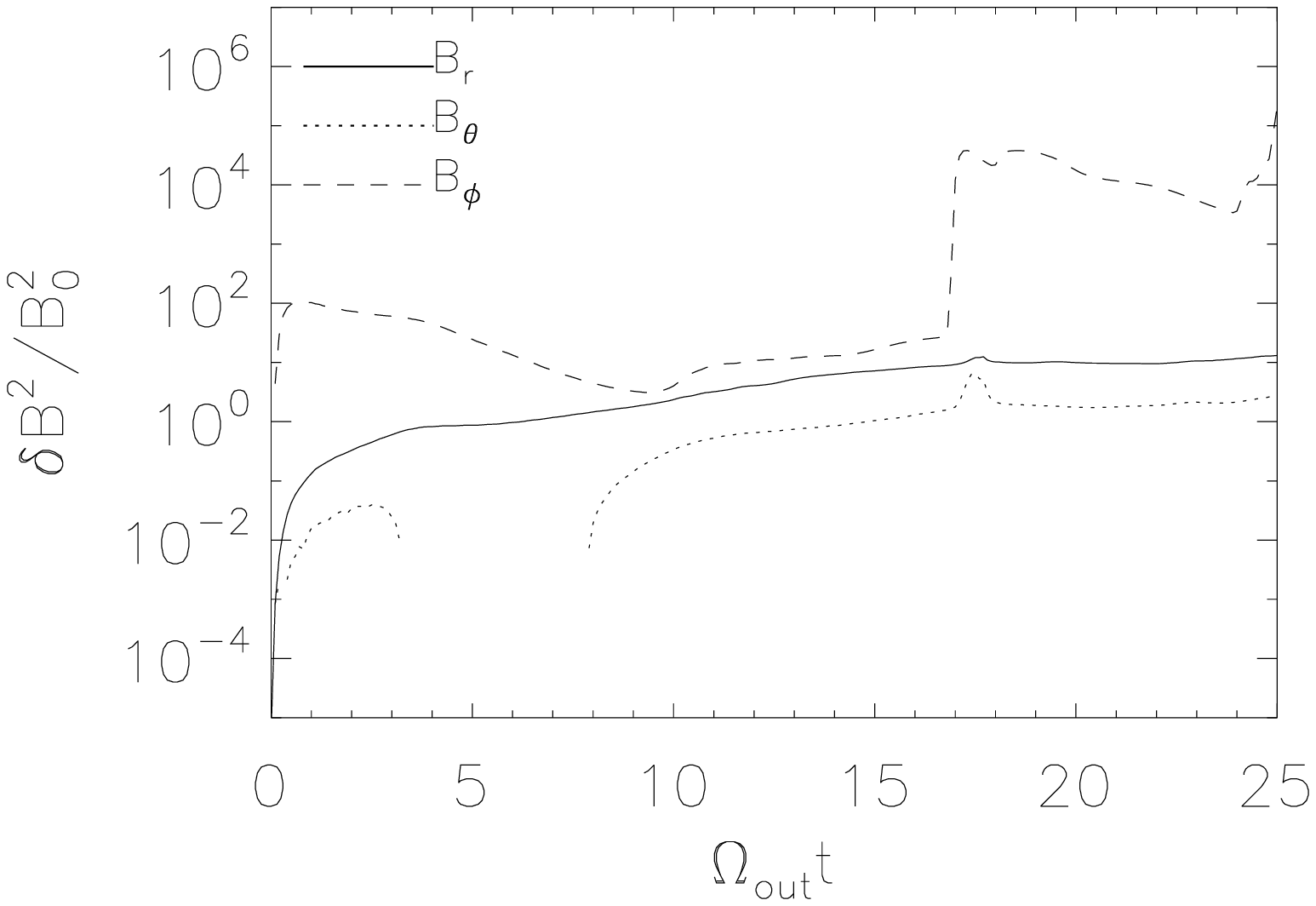}

\end{center}
\caption{Change of volume averaged magnetic energy in the $r$, $\theta$
and $\phi$ components of the magnetic field normalized to the initial magnetic energy in
a shell from $r_{\rm out}/40$ to $r_{\rm out}$ as a function of time
in Model H1. In this model, magnetic field is amplified by the MRI. The
growth of each component of the magnetic field is defined as $\delta
B^2_{r,\theta,\phi} = ({\langle
{B^2}_{r,\theta,\phi}\rangle}_V-B^2_{r,\theta,\phi,0})$, where
$\langle\rangle_V$ denotes volume average and $B_0$is the initial
field strength.}
\end{figure}

\begin{figure}
\begin{center}
\includegraphics[width=8.5cm]{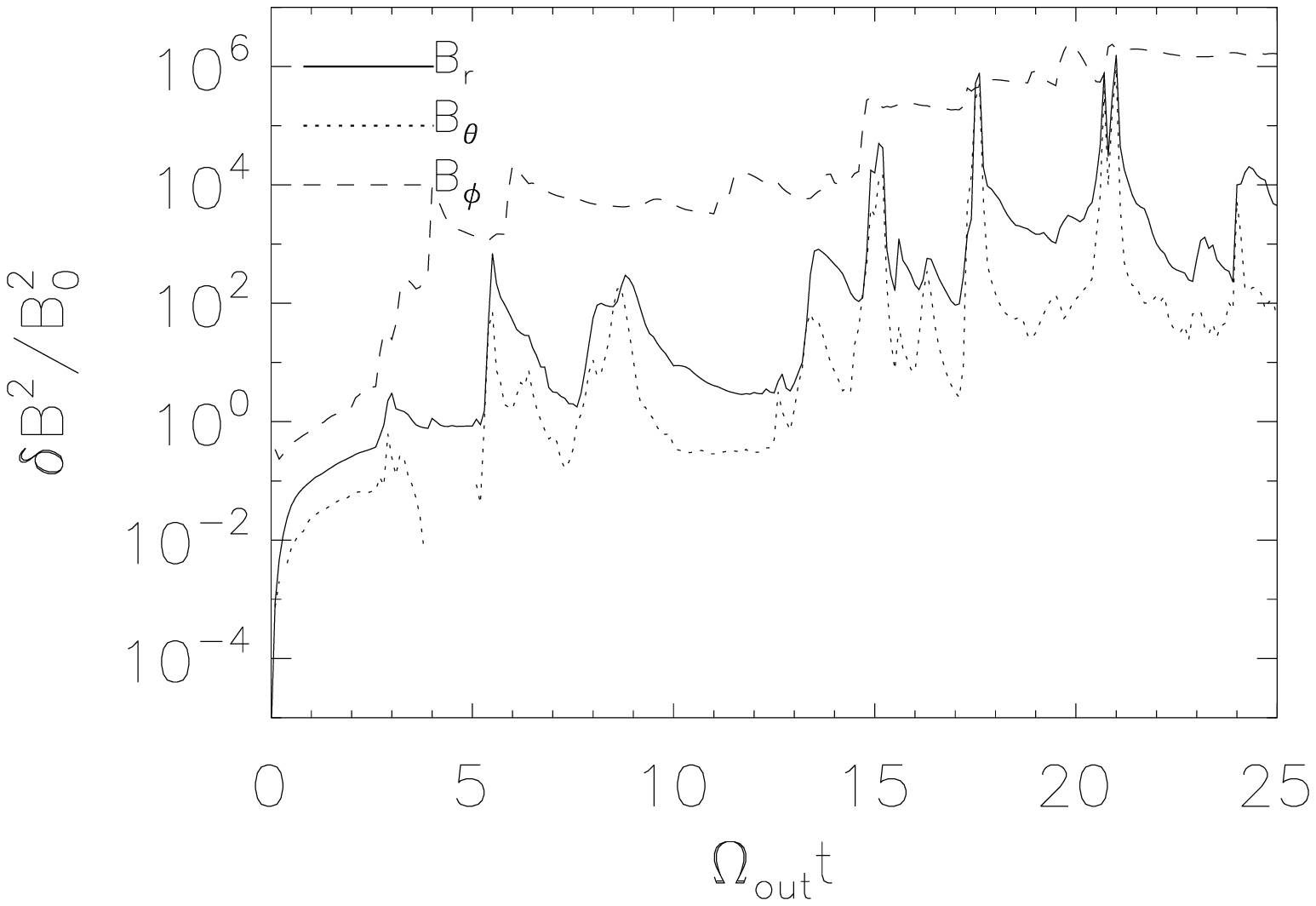}

\end{center}
\caption{Change of volume averaged magnetic energy  in the $r$, $\theta$
and $\phi$ components of the magnetic field normalized to the initial magnetic energy in
a shell from $r_{\rm out}/40$ to $r_{\rm out}$ as a function of time
in Model H2. In this model, magnetic field is amplified by both the MTI
and MRI. The growth of each component of the magnetic field is
defined as $\delta B^2_{r,\theta,\phi} = ({\langle
{B^2}_{r,\theta,\phi}\rangle}_V-B^2_{r,\theta,\phi,0})$, where
$\langle\rangle_V$ denotes volume average and $B_0$ is the initial
field strength.}
\end{figure}

\begin{figure}
\begin{center}
\includegraphics[width=8.5cm]{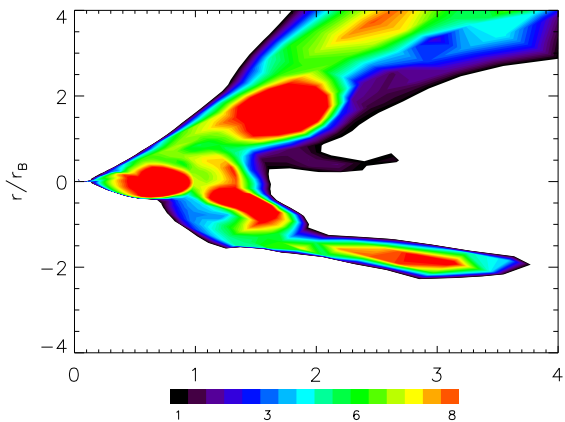}

\end{center}
\caption{Contour of $Re=\lambda_{MRI}/\Delta x $ at time $t=16.5$,
where $\Delta x$ is the grid size.  It is clear that the wavelength
of the fastest growing mode of the MRI is well resolved at time
$t=16.5$.}
\end{figure}

We investigate field amplification by the MRI in model H1. In this
model, we turn off thermal conduction so the MTI does not exist, and
the magnetic field amplification must be due to motions driven by
the MRI. Figure 3 shows the time evolution of the magnetic energy.
Initially ($t<16.5$), the field is very weak, and at our resolution
the fastest growing mode of the MRI is not resolved, so that field
amplification occurs primarily by flux freezing. The initial ($t<1$)
rapid amplification of the azimuthal component of the magnetic field
is due to shear. At $t=16.5$, the magnetic field strength is
amplified by flux freezing to a high enough value that the MRI is
resolved, and the field undergoes substantial and rapid
amplification by the MRI. To illustrate this point, we define a
parameter $Re=\lambda_{MRI}/\Delta x $, where $\Delta x$ is the grid
size, which measure how well the MRI is resolved on our numerical
grid. Figure 5 shows contours of $Re$ at $t=16.5$. From this figure,
it is clear that unstable wavelengths of the MRI are well resolved
at that time.  The MRI amplifies the field rapidly, and shortly
after $t=16.5$ the MRI saturates and the field begins to grow
algebraically.  Saturation of the MRI can be understood as follows.
The condition for MRI to operate is $k^2 v_a^2+d\Omega^2/d{\rm ln}r<
0$. After $t=18$, we find $k^2 v_a^2+d\Omega^2/d{\rm ln}r \sim 0$ in
the region where the MRI growth wave length can be resolved, so the
growth rate of MRI decreases significantly. Using parameters
$F_{r,r}$, $F_{\theta,r}$, and $F_{\phi,r}$ to represent the
magnetic field amplification factor of the $r$, $\theta$ and $\phi$
components of the magnetic field, respectively, we find
\begin{equation}
F_{r,r}=B_{r,f}^2/B_0^2 \sim 10,
\end{equation}
\begin{equation}
F_{\theta,r}=B_{\theta,f}^2/B_0^2 \sim 10,
\end{equation}
\begin{equation}
F_{\phi,r}=B_{\phi,f}^2/B_0^2 \sim 10^4,
\end{equation}
where $B_{r,f}$, $B_{\theta, f}$ and $B_{\phi,f}$ correspond to the
$r$, $\theta$ and $\phi$ components of the magnetic field at the end
of the simulation, respectively. The toroidal component is most
significantly amplified as expected.

In model H2, we turn on thermal conduction so both the MTI and MRI
coexist. Figure 4 shows the time evolution of the magnetic energy.
Initially ($t<3$), there is only the MTI. This is because initially
the field is too weak for unstable wavelengths of the MRI to be
resolved. At $t=3$ the magnetic field has been amplified by the MTI
enough that unstable modes of the MRI are resolved, and the MRI
begins to drive motions. After $t=3$, the field begins to be
amplified exponentially by both the MTI and MRI, until at $t \sim
20$ both instabilities saturate and the growth of the field becomes
algebraic. Using parameters $F_{r}$, $F_{\theta}$ and $F_{\phi}$ to
represent the amplification factor of the $r$, $\theta$ and $\phi$
components of the magnetic energy, respectively, we find
\begin{equation}
F_{r}=B_{r,f}^2/B_0^2 \sim 10^3,
\end{equation}
\begin{equation}
F_{\theta}=B_{\theta,f}^2/B_0^2 \sim 100,
\end{equation}
\begin{equation}
F_{\phi}=B_{\phi,f}^2/B_0^2 \sim 10^6,
\end{equation}
where $B_{r,f}$, $B_{\theta, f}$ and $B_{\phi,f}$ correspond to the
$r$, $\theta$ and $\phi$ components of the magnetic field at the end
of the simulation, respectively.

Comparing models L2, H1 and H2, we find,
\begin{equation}
F_r \sim F_{r, t} \cdot F_{r,r},
\end{equation}
\begin{equation}
F_\theta \sim F_{\theta, t} \cdot F_{\theta,r},
\end{equation}
\begin{equation}
F_\phi \sim F_{\phi, t} \cdot F_{\phi,r}.
\end{equation}
Let us first consider the amplification of the radial component of
the magnetic field. If the MTI alone amplifies the radial component of
the magnetic energy by a factor of $F_{r,t}$ (model L2), and the MRI alone
by a factor of $F_{r,r}$ (model H1), then when the MTI and MRI coexist the
radial component of the magnetic field can by amplified by a factor
of $F_{r,t} \cdot F_{r,r}$ (model H2). The amplification factors
for the other two
components of the magnetic energy behave similarly. We therefore
conclude that in the nonlinear regime the
MTI and MRI amplify the field independently, most likely
because the
MTI is buoyancy instability driven by radial gradients in the temperature,
while the MRI is a dynamical instability driven by the radial gradient of
the orbital frequency.  Provided that the
magnetic field is weak and the plasma is dilute, thermal conduction
from hotter to cooler regions will make the plasma buoyantly unstable,
regardless of whether the flow also contains turbulent eddies driven by the
MRI. On the other hand,
provided that the magnetic field is weak and
differential rotation exists, the MRI can operate regardless of the
presence of convective eddies driven by the MTI.

\subsection{The role of MTI in angular momentum transfer}

The MRI is now believed to be the most promising mechanism of
angular momentum transport in fully ionized accretion flows. It is
interesting to investigate whether the MTI can contribute to angular
momentum transfer in hot, diffuse accretion flows.

We first compare the angular momentum profiles in models L1 and L2
at the end of both calculations. The flow in model L1 is laminar,
while there is (only) MTI in model L2.  Angular momentum profiles at
the end of the simulations are shown in Figure 6.  It is clear that
the angular momentum distribution of model L1 is almost the same as
the initial distribution. This is because the flow in model L1 is
laminar, and the magnetic field is very weak: therefore both the
Reynolds and magnetic stresses are very small and there is almost no
angular momentum transfer.  It is also cleat that in model L2,
angular momentum is transferred from the inner to the outer regions.
In order to investigate the mechanism for the angular momentum
transport in this case, we plot the Maxwell and Reynolds stresses in
Figure 7. The dotted and solid lines correspond to the Maxwell
stress ($-B_rB_{\phi}/4\pi$) in models L1 and L2, respectively.
Because of magnetic field amplification by the MTI, the Maxwell
stress in model L2 is much larger than that in model L1. The
triangles and squares denote the Reynolds stress in models L1 and L2
respectively. Due to the presence of motions driven by the MTI in
model L2, the Reynolds stress is three orders of magnitude higher
than in model L1.  Because the magnetic field is still very weak in
model L2, the Maxwell stress is much smaller than the Reynolds
stress. We conclude the angular momentum transport in model L2 is
dominated by the Reynolds stress. We note that although the MTI
induced Reynolds stress does transfer angular momentum, the effect
is quite low, so that there is only a small change of the angular
momentum profile compared to the initial conditions.

Next we compare the angular momentum distribution in models H1 and
H2 at late times in Figure 8. Both profiles are quite similar.
Recall that both models are unstable to the MRI, while only H2 is
unstable to the MTI.  The similarity in the angular momentum
profiles indicates that the role of the MTI in angular momentum
transport is less important compared to the MRI.
\begin{figure}
\begin{center}
\includegraphics[width=8.5cm]{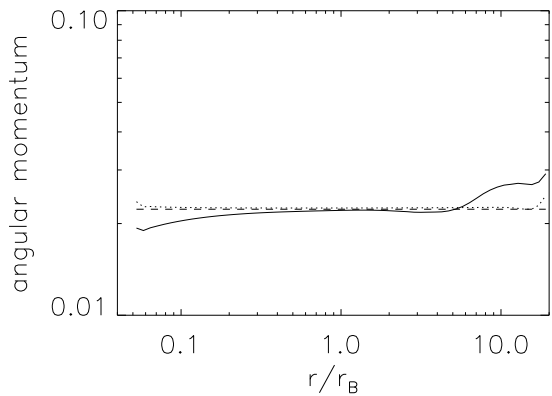}
\end{center}
\caption{Radial profiles of specific angular momentum for models L1
(dotted line) and L2 (solid line). The dashed line shows the initial
angular momentum distribution.}
\end{figure}

\begin{figure}
\begin{center}
\includegraphics[width=8.5cm]{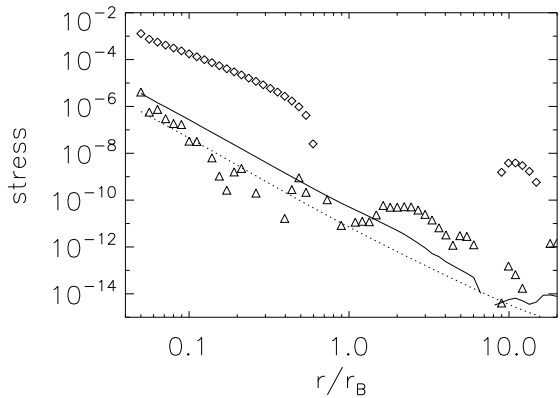}
\end{center}
\caption{Radial profiles of stress. The dotted and solid lines
correspond to the Maxwell stress in models L1 and L2, respectively.
The triangle and square are Reynolds stress in models L1 and L2,
respectively.}
\end{figure}

\begin{figure}
\begin{center}
\includegraphics[width=8.5cm]{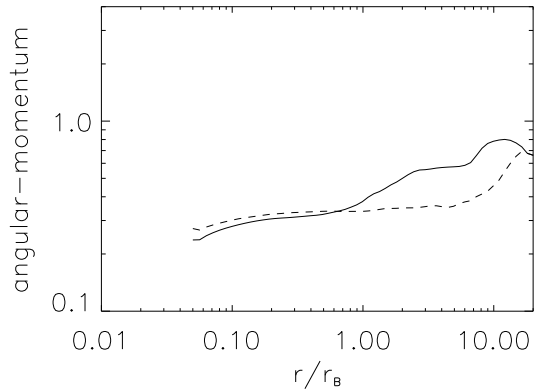}
\end{center}
\caption{Radial profiles of specific angular momentum for models H1
(solid line) and H2 (dashed line).}
\end{figure}

\subsection{Effects of MTI on the structure of the accretion flow}

Figure 9 shows the time-averaged density (upper-left panel),
temperature (upper-right) and mass accretion rate (lower panel) for
models H1 and H2. In the upper panels, the solid and dashed lines
correspond to models H1 and H2, respectively. In the
lower panel, the solid, dotted and dashed lines correspond to the
net accretion rate, inflow rate and outflow rate, respectively. The
black and red lines correspond to models H1 and H2, respectively. As
shown in the upper-right panel, the radial profile of temperature in
model H2 (with MTI) is flatter than that in model H1. This is likely
because thermal conduction transports energy from the inner
to the outer regions of the flow. From the lower panel, we can see
that the inflow rate decreases due to the effects of thermal
conduction. This is consistent with Johnson \& Quataert (2007). The
decrease is caused by the increased heating of the outer regions
by thermal conduction.
\begin{figure}
\begin{center}
\includegraphics[width=8.5cm]{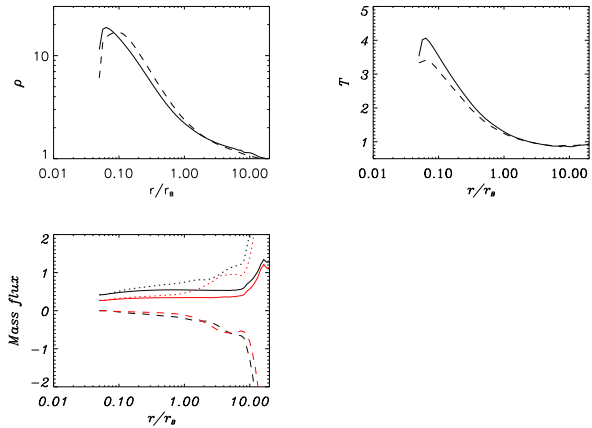}
\end{center}
\caption{Radial profiles of time-averaged variables in models H1 and
H2. Upper-left and upper-right panels correspond to the radial
profiles of density and temperature, respectively. In the upper
panels, solid and dashed lines correspond to models H1 and H2,
respectively. In the lower panel, the solid, dotted and dashed lines
correspond to the net accretion rate, inflow rate and outflow rate,
respectively. The black and red lines correspond to models H1 and
H2, respectively.}
\end{figure}

\section{Summary and discussion}

In extremely low accretion rate systems, the plasma is very dilute,
the collisional mean-free path of electrons is much greater than their
Larmor radius; thus thermal conduction is anisotropic and along
magnetic field lines. Anisotropic thermal conduction can
fundamentally alter the Schwarzschild convective instability
criterion. Convection sets in when temperature (as opposed to
entropy) increases in the direction of gravity.
Convective instability in this regime is referred to as the
magnetothermal instability (MTI).
The MTI can amplify the magnetic field and align field lines with the
temperature gradient (i.e., the radial direction). If the flow has
angular momentum, the MRI may also exist. We have investigated the possible
interaction between the MTI and MRI by performing two-dimensional
magnetohydrodynamical (MHD) simulations of nonradiative rotating
accretion flows with anisotropic thermal conduction.

We have presented the results from four models, with focus on the
amplification of magnetic energy by the combination of MTI and MRI.
As a control case, we computed a model (L1) in which neither the MTI
nor the MRI was present, so that the magnetic field is amplified
only by flux freezing and geometrical compression in the flow. We
find that field amplification via flux freezing in this case is very
small.  Next, we investigated the effects of the MTI on magnetic
field amplification in model L2. We find that the MTI can amplify
magnetic field significantly, which is consistent with the result
obtained in previous work.  Finally we considered the evolution of
rotating flows that are unstable to the MRI.  In model H1 (no
thermal conduction), only the MRI is present.  In model H2, both the
MTI and MRI coexist. By comparing the structure and evolution of
these different models, we find that if the MTI alone can amplify
the magnetic energy by a factor of $F_t$, and the MRI alone by a
factor of $F_r$, then when the MTI and MRI are both present, the
magnetic energy can be amplified by a factor of $F_t\cdot F_r$.
Thus, we conclude that the MTI and MRI operate independently, and
that there is little feedback on the amplification factor in the
nonlinear regime.  Because the MTI and MRI are driven by different
properties of the flow (temperature gradient in the case of the
former, and angular velocity gradient in the case of the latter),
the presence of one may not strongly affect the presence of the
others. Thus, anisotropic thermal conduction can make the plasma
buoyantly unstable, regardless of whether or not turbulent eddies
driven by the MRI are present.  Similarly, an unstable rotation
profile can result in the MRI, regardless of whether or not there
are convective eddies induced by the MTI.

In addition to the interplay between the MTI and MRI, we have also
investigated the effects of MTI on angular momentum transport.  We
find that the MTI enhances angular momentum transport simply because
it can amplify magnetic field (thus enhancing the Maxwell stress)
and induce radial motions (thus enhancing the Reynolds stress).  But
we note that the effects of the MTI on angular momentum transfer are
small. Finally, we find that thermal conduction does make the slope
of the temperature smaller due to the outward transport of energy,
and the increase of the temperature in the outer regions decreases
the mass accretion rate.

Lastly, we would like to mention that in a weakly collisional
accretion flow, the ion mean-free path can be much greater than its
Larmor radius; thus the pressure tensor is anisotropic. In this
case, the growth rate of the MRI can increase dramatically at small
wavenumbers compared to the MRI in ideal MHD (Quataert, Dorland \&
Hammett 2002; Sharma, Hammett \& Quataert 2003).   In this regime,
the viscous stress tensor is anisotropic, and Balbus (2004; see also
Islam \& Balbus 2005) has shown that when anisotropic viscosity is
included, the flow is subject to the magnetoviscous instability
(MVI). The MVI may also amplify magnetic field significantly.  An
interesting project for the future is to investigate the effects of
the MVI on hot accretion flows.

\section{ACKNOWLEDGMENTS}

We thank P. Sharma for sending us his code. This work was supported
in part by the Natural Science Foundation of China (grants 10773024,
10833002, 10821302, and 10825314), the National Basic Research
Program of China (973 Program 2009CB824800), and the CAS/SAFEA
International Partnership Program for Creative Research Teams.



\begin{thebibliography}{}
\bibitem[]{} Balbus S. A., 2000, ApJ, 534, 420
\bibitem[]{} Balbus S. A., 2001, ApJ, 562, 909
\bibitem[]{} Balbus S. A., 2004, ApJ, 616, 857
\bibitem[]{} Balbus S. A., Hawley J. F., 1991, ApJ, 376, 214
\bibitem[]{} Balbus S. A., Hawley J. F., 1998, Rev. Mod. Phys., 70,1
\bibitem[]{} Igumenshchev I. V., Abramowicz M. A., 1999, MNRAS,
303,309
\bibitem[]{} Igumenshchev I. V., Abramowicz M. A., 2000, ApJS,
130,463
\bibitem[]{} Islam T., Balbus S., 2005, ApJ, 633, 328
\bibitem[]{} Johnson B. M., Quataert E., 2007, ApJ, 660, 1273
\bibitem[]{} Narayan R., 2005, Ap\&SS, 300, 177
\bibitem[]{} Narayan R., McClintock J.E., 2008, New Astron. Rev., 51, 733
\bibitem[]{} Narayan R., Yi I., 1994, ApJ, 428, L13
\bibitem[]{} Narayan R., Yi I., 1995, ApJ, 452, 710
\bibitem[]{} Parrish I. J., Stone J. M., 2005, ApJ, 633, 334
\bibitem[]{} Parrish I. J., Stone J. M., 2007, ApJ, 664, 135
\bibitem[]{} Parrish I. J., Stone J. M., Lemaster N., 2008, ApJ, 688,
905
\bibitem[]{} Quataert E., 2004, ApJ, 613, 322
\bibitem[]{} Quataert E., Dorland W., Hammett G. W., 2002, ApJ, 577,
524
\bibitem[]{} Sharma P., Hammett G. W., 2007, J. Comp. Phys.,
227,123
\bibitem[]{} Sharma P., Hammett G. W., Quataert E., 2003, ApJ, 596,
1121
\bibitem[]{} Sharma P., Quataert E., Stone J. M., 2008, MNRAS, 389,
1815
\bibitem[]{} Stone J.M., Norman M. L., 1992a, ApJS, 80, 753
\bibitem[]{} Stone J.M., Norman M. L., 1992b, ApJS, 80, 791
\bibitem[]{} Stone J.M., Pringle J. E., Begelman M. C., 1999, MNRAS,
310, 1002
\bibitem[]{} Tanaka T., Menou K., 2006, ApJ, 649, 345
\bibitem[]{} Yuan F., 2007, in Luis C. Ho., Jian-Min Wang, eds, ASP
conf. Ser., Vol.373, The Central Engine of Active Galactic Nuclei.
Astron. Soc. Pac., San Francisco, p.95
\bibitem[]{} Yuan F., Bu D. F., 2010, MNRAS, 408, 1051
\end{thebibliography}
\end{document}